\documentclass[fleqn,11pt]{article}
\usepackage{a4}
\usepackage{graphicx}
\usepackage{psfrag}
\usepackage{beton}
\usepackage{concrete}
\usepackage{dsfont}

\newtheorem{definition}{Definition}
\newtheorem{lemma}[definition]{Lemma} 
\newtheorem{observation}[definition]{Observation} 
\newtheorem{theorem}[definition]{Theorem}

\newsavebox{\proofsymbol}   
\savebox{\proofsymbol}                         
{
    \begin{picture}(10,10)                       
    \put(0,0){\framebox(9,9){}}                 
    \put(0,3){\framebox(6,6){}}                  
    \end{picture}                               
} 

\newcommand{\proof}{\noindent{\bf Proof.} \hspace{1mm}}
\newcommand{\qed}{\hfill\usebox{\proofsymbol}}

\newcommand{\R}{\mathds{R}}
\newcommand{\lex}{\rm lex}

\begin{document}

\title{Ranking Unit Squares with Few Visibilities} 
\author{Bernd G\"artner\thanks{
Institute of Theoretical Computer Science, 
ETH Z{\"u}rich, CAB G32.2, CH-8092 Z{\"u}rich, Switzerland. Email:
gaertner@inf.ethz.ch}}
\date{August 4, 2008} 
\maketitle

\begin{abstract}
  Given a set of $n$ unit squares in the plane, the goal is to rank
  them in space in such a way that only few squares see each other
  vertically.  We prove that ranking the squares according to the
  lexicographic order of their centers results in at most $3n-7$
  pairwise visibilities for $n\geq 4$.  We also show that this bound is
  best possible, by exhibiting a set of $n$ squares with at least
  $3n-7$ pairwise visibilities under any ranking.
\end{abstract}

\section{Problem statement}
The unit square with center $c\in\R^2$ is the set
\[S(c) = \{p\in\R^2\mid \|p-c\|_{\infty}\leq 1\}.\] Given a set ${\cal
  S}$ of $n$ unit squares (simply called squares in the sequel), a
\emph{ranking} of ${\cal S}$ is a sequence
$\rho=(S_1,S_2,\ldots,S_n)$ such that ${\cal
  S}=\{S_1,S_2,\ldots,S_n\}$. For $i<k$, $S_i$ \emph{sees}
$S_k$ under $\rho$ if there exists a point $p\in\R^2$ such that
\[
\begin{array}{rcll}
p&\in& S_i\cap S_k,\\
p&\notin& S_j, & i<j<k.
\end{array}
\]

The graph on ${\cal S}$ formed by all pairs $\{S_i,S_k\}$ such that
$S_i$ sees $S_k$ under $\rho$ is called the \emph{visibility graph}
of $\rho$ and will be denoted by $G(\rho)$. 

The goal is to find a ranking $\rho$ such that $G(\rho)$ has as
few edges as possible. We do not know how to find the best ranking for
a given set ${\cal S}$, but we prove that there is always a ranking 
$\rho$ under which $G(\rho)$ has no more than $3n-7$ edges. For
some sets ${\cal S}$, this is the best bound that can be achieved.

This research is motivated by similar questions for intervals in
$\R^1$ \cite{intervals}.

\section{Main result}
Given a ranking $\rho=(S_1,S_2,\ldots,S_n)$, the center of square
$S_i$ will be denoted by $c_i=(x_i,y_i), i=1,2,\ldots,n$.  We will
repeatedly use the following simple fact.

\begin{observation}\label{obs:help}
  Let $\rho=(S_1,S_2,\ldots,S_n)$ be a ranking, and suppose that
  there are centers $c_i,c_j,c_k$ with $i<j<k$, such that $c_j$ is
  contained in the (axis-parallel) rectangle spanned by $c_i$ and
  $c_k$. Then $\{S_i,S_k\}$ is not an edge of~$G(\rho)$.
\end{observation}

\proof 
Center $c_j$ being contained in the rectangle spanned by $c_i$
and $c_k$ is easily seen to be equivalent to $S_j\supseteq S_i\cap S_k$.
It follows that $S_i$ does not see $S_k$ under $\rho$.  
\qed

\begin{figure}[htb]
\begin{center}
\includegraphics[width=12cm]{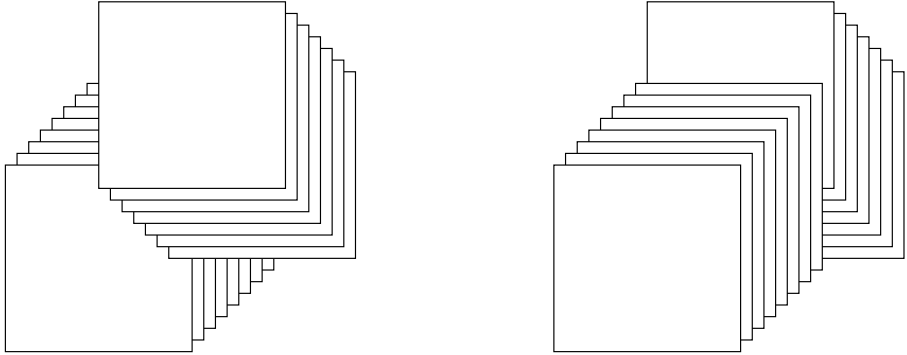}
\end{center}
\caption{Upward view on a ranked set of squares. Left: All squares in the 
lower half see all squares in the
upper half. Right: the lexicographic ranking incurs only linearly many 
visibilities.\label{fig:n2}}
\end{figure}

The size (number of edges) of the visibility graph may be
$\Theta(n^2)$ for a ``bad'' ranking, see Figure~\ref{fig:n2} (left).
The right part of the figure depicts the lexicographic ranking.
According to the next lemma, this ranking always results in $O(n)$
visibilities.

\begin{lemma}\label{lem:main}
  Let $\rho_{\lex}=(S_1,S_2,\ldots,S_n)$ be the lexicographic ranking,
  meaning that $i<j$ if and only if the center of $S_i$ is
  lexicographically smaller than the center of $S_j$. Then $G(\rho_{\lex})$
  is a planar graph. More precisely, the straight-line embedding of
  $G(\rho_{\lex})$ obtained by mapping each square $S_i$ to its center
  $c_i\in\R^2$ is a plane graph.
\end{lemma}

\proof We show that if two segments $\overline{c_ic_j}$ and
$\overline{c_kc_{\ell}}$ cross, then $G(\rho_{\lex})$ contains at most
one of the edges $\{S_i,S_j\}$ and $\{S_k,S_{\ell}\}$. 

Since deletion of other squares can only add visibilities between the
four squares involved in the crossing, we may w.l.o.g.\
assume that ${\cal S}=\{S_1,S_2,S_3,S_4\}$. By 
lexicographic order of the centers, there are only two
cases.\smallskip

\emph{Case (a): Segments $\overline{c_1c_3}$ and $\overline{c_2c_4}$
  cross.} If $y_2$ is between $y_1$ and $y_3$, we get that
$c_2$ is in the rectangle spanned by $c_1$ and $c_3$, so
$\{S_1,S_3\}$ is not an edge of $G(\rho_{\lex})$ by 
Observation~\ref{obs:help}.

If $y_2$ is not between $y_1$ and $y_3$, we either have
$y_2<\min(y_1,y_3)$ and thus $y_4>y_3$ (otherwise, there
would be no crossing), or $y_2>\max(y_1,y_3)$ and thus $y_4 <
y_3$. In both cases, $y_3$ is between $y_2$ and $y_4$ which
means that $c_3$ is in the rectangle spanned by $c_2$ and $c_4$. This
in turn shows that $\{S_2,S_4\}$ is not an edge of $G(\rho_{\lex})$.
\smallskip

\emph{Case (b): Segments $\overline{c_1c_4}$ and $\overline{c_2c_3}$
  cross.} An easy case occurs if one of
$c_2$ and $c_3$ is in the rectangle spanned by $c_1$ and
$c_4$, since Observation~\ref{obs:help} then implies that
$\{S_1,S_4\}$ is not an edge.

Otherwise, we have 
\begin{equation}\label{eq:visa}
y_2<\min(y_1,y_4)\leq y_4
\end{equation}
and thus 
\begin{equation}\label{eq:visb}
y_3>\max(y_1,y_4)\geq y_1
\end{equation}
(because of the crossing), or $y_3<\min(y_1,y_4)$ and
$y_2>\max(y_1,y_4)$. Let us only treat the first case; the second one
is symmetric under exchange of indices~$2$ and~$3$.

We now show that every point in $S_1\cap S_4$ is in $S_2\cup
S_3$, given that $S_2\cap S_3\neq\emptyset$. Therefore, if
$\{S_2,S_3\}$ is an edge of $G(\rho_{\lex})$, then
$\{S_1,S_4\}$ is not an edge.

Let $p=(x,y)$ be any point in $S_1\cap S_4$. 
From $x_4-1\leq x \leq x_1+1$
and lexicographic order it follows that
\begin{equation}
x_2-1, x_3-1 \leq x \leq x_2+1, x_3+1.\label{eq:vis1}
\end{equation}
Using $y_4-1 \leq y \leq y_1+1$ together with (\ref{eq:visa}) and
(\ref{eq:visb}), we also conclude that
\begin{equation}
y_2-1 \leq y \leq y_3+1.\label{eq:vis2}
\end{equation}
If $y$ in addition satisfies $y\leq y_2+1$, (\ref{eq:vis1}) and
(\ref{eq:vis2}) imply that $p\in S_2$. But if $y>y_2+1$, we can
use the assumption $S_2\cap S_3\neq\emptyset$ to conclude that 
$y_3-y_2\leq 2$ and hence $y>y_3-1$. With (\ref{eq:vis1}) and
(\ref{eq:vis2}), we then get $p\in S_3$.
\qed

Now we are ready to prove our main theorem.

\begin{theorem}\label{thm:main}
If $n\geq 4$, $G(\rho_{\lex})$ has at most $3n-7$ edges.
\end{theorem}

\proof An upper bound of $3n-6$ already follows from Lemma
\ref{lem:main}. In order for this bound to be tight, the outer face of
$G(\rho_{\lex})$'s straight-line embedding would have to be a
triangle $\Delta$ spanned by three centers
$c_i, c_j,c_k$, $i<j<k$, and with
all other centers inside $\Delta$. Indeed, for $n=3$ this is possible,
but for $n\geq 4$, we get a contradiction: Let $c=(x,y)$ be any center
distinct from $c_i,c_j,c_k$. From $c\in\Delta$, it easily follows that
$c_i< c <c_k$ (comparison being lexicographically). 

If $y$ is between $y_i$ and $y_k$, $c$ is in the rectangle spanned by
$c_i$ and $c_k$, so $\{S_i,S_k\}$ can't be an edge of
$G(\rho_{\lex})$ by Observation \ref{obs:help}.

If $y$ is not between $y_i$ and $y_k$, then $y$ must be between $y_i$ and
$y_j$, \emph{and} between $y_j$ and $y_k$. Depending on whether
$c_i < c < c_j$ or $c_j < c < c_k$, we get that either 
$\{S_i,S_j\}$ or $\{S_j,S_k\}$ is not an edge of $G(\rho_{\lex})$.
\qed

\section{Lower Bound}
\begin{figure}[htb]
\begin{center}
\includegraphics[width=5cm]{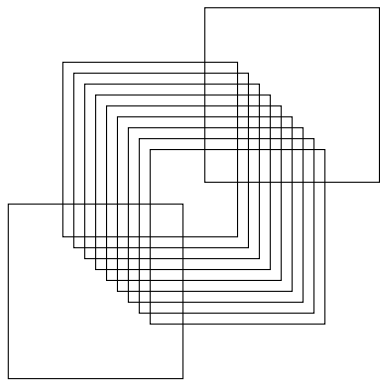}
\end{center}
\caption{A set of $n$ unit squares such that the visibility graph
  $G(\rho)$ has at least $3n-7$ edges for every ranking $\rho$: a
  square from the middle bunch of $n-2$ squares always sees the next
  square above it in the bunch (these are $n-3$ visibilities), and it
  sees (or is seen by) both of the two special squares ($2n-4$
  visibilities).\label{fig:best}}
\end{figure}

The bound of $3n-7$ derived in Theorem \ref{thm:main} is best possible
in the worst case, not only under the lexicographic ranking, but under
\emph{every} ranking. For a proof by picture see Figure \ref{fig:best}.

\section*{Acknowledgment}
The problem of ranking unit squares was given to me as an
undergraduate student by Emo Welzl in early 1989.  After some initial
programming tasks, this was the first actual research assignment
during my time as ``Forschungstutor'' (research assistant) with Emo.

My initial manuscript from 1989, typeset in bumpy \LaTeX\ and
containing manually drawn figures, is lost. But throughout the almost
twenty years that have passed since then, I could never forget this
first (simple) result of mine. 

While writing it up now, many memories of all these years have come
back to me with surprising strength. I thank Emo for bringing the
problem to my attention, and for so much more.
\bibliographystyle{plain} \bibliography{biblio}

\end{document}